\documentclass[prd,nofootinbib,showpacs,superscriptaddress,preprint]{revtex4}
\usepackage[T1]{fontenc}
\usepackage{amsmath,amssymb}
\usepackage{epsfig}
\usepackage{dcolumn}
\usepackage{mathalfa}
\usepackage[usenames]{color}
\usepackage{slashed}
\usepackage[colorlinks,citecolor=blue]{hyperref}
\usepackage{pdfpages}
\usepackage{float}
\usepackage{graphicx}
\usepackage{xcolor}
\usepackage[font=footnotesize]{caption}
\graphicspath{{fignh/}{figih/}}
\let\revappendix\appendix

\def\bea{\begin{eqnarray}}
	\def\eea{\end{eqnarray}}

\begin{document}
	
	\title{Leptogenesis and dark matter in minimal inverse seesaw using $A_4$ modular symmetry}
	\author{Jotin Gogoi}
	\email{jotingo@tezu.ernet.in}
	\author{Lavina Sarma}
	\email{lavina@tezu.ernet.in}
	\author{Mrinal Kumar Das}
	\email{mkdas@tezu.ernet.in}
	\affiliation{Department of Physics, Tezpur University, Tezpur 784028, India}
	
	\begin{abstract}
		
		In this paper we have studied neutrino masses and mixings by adding a scalar triplet $\eta$ to the particle content of minimal Inverse seesaw. We have realised this extension of minimal inverse seesaw by implementing an isomorphic modular group $\Gamma(3)$ and a non-abelian discrete symmetry group $A_4$. We have also used $Z_3$ symmetry group to restrain certain interaction terms in the lagrangian of the model. We have studied baryon asymmetry of the universe, neutrinoless double-beta decay and dark matter in our work. In order to check the consistency of our model with various experimental constraints, we have therefore calculated effective mass, relic density and baryogenesis via leptogenesis. Interestingly, we have found our model quite compatible with the experimental bounds and is also successful in producing the neutrino masses and mixings in the 3$\sigma$ range.

		\vspace{3cm}

	\end{abstract}

	\maketitle
	
		\section{Introduction}
		
		With the discovery of the Higgs particle at the LHC in 2012,  a very important revolution was marked in the field of particle physics \cite{doi:10.1063/1.4727988}. This discovery, apart from providing the missing piece to the Standard Model (SM), paved the way for a better understanding of the fundamental particles and through Higgs mechanism the process responsible for generating the masses of quarks and leptons could be explained. But this grand success was not devoid of shortcomings. There are many challenging issues in cosmology, astrophysics, particle physics etc. which are still not addressed and the answers to these phenomena are yet to be found. Some of these prominent challenges are neutrino oscillations and generation of their tiny masses, lepton number violating processes (LNV), baryon asymmetry of the universe (BAU), dark matter, dark energy etc. The Standard Model of particle physics, recognised as one of the most successful theories in the field, fails to provide a proper explanation of these challenges. As a result, the concept of Beyond the Standard Model (BSM) frameworks have come into the picture. And people believe that a proper understanding with a suitable explanation of these challenging issues can be found in this BSM paradigm.
		
		One of the prime ingredients that motivates physicists to go for BSM theories is associated with mass of the neutrinos. With the discovery of neutrino oscillation \cite{Bellini:2013wra}, it became obvious that neutrinos should have tiny mass and mixing between them. The experimental findings made by Super-Kamiokande \cite{KamLAND:2008dgz,Super-Kamiokande:2001bfk} and Sudbury Neutrino Observatories \cite{SNO:2002hgz} gave the final confirmation to this idea. But due to the absence of right-handed neutrinos it remained massless within the Standard Model. As a result, to generate this tiny mass of neutrinos an extension of SM was inevitable. Consequently a new theory that could incorporate and provide an elegant explanation to these questions came in the form of Seesaw mechanism. A detailed review about seesaw mechanism can be found in the literature \cite{Babu:1988qv,Chang:1985en,Mohapatra:1987fs}. This mechanism is broadly classified into Type I \cite{Albright:2004kb,Mohapatra:2004vr,King:2003jb}, Type II \cite{Rodejohann:2004qh}, Type III \cite{King:2014nza}, Inverse seesaw etc. In recent times, Inverse seesaw has become more popular in model building. In this framework the mass of right-handed neutrinos can be brought down to TeV scale. This peculiar property enhances the possibility of detecting these particles at the LHC and future experiments. For a detailed study on Inverse seesaw one can refer to \cite{Abada:2018qok,Deppisch:2004fa,Dev:2009aw}.

		It is now an established fact that there is an asymmetry between matter and anti-matter in the universe. Cosmological observations indicate that the number of baryons in the universe is unequal to the number of anti-baryons. This difference in number between baryons and anti-baryons is termed as Baryon Asymmetry of the Universe (BAU) \cite{Barbieri:1999ma,Kolb:1979ui,Pilaftsis:2003gt}. At the beginning, as evident from various considerations, the universe started with an equal number of both the types of particles i.e. baryons and anti-baryons. As such the asymmetry observed in the universe occured in much later times and must have been generated through a dynamical process called baryogenesis. As per Planck data, the value of this asymmetry is found to be \cite{ParticleDataGroup:2018ovx}
		\begin{equation}
			\eta_{B}=(6.04\pm 0.08)\times 10^{-10}
			\label{seceq1}
		\end{equation}
	
	As proposed by Sakharov, three important conditions are necessary for baryogenesis: Baryon number violation (B), C and CP violation , interactions out of thermal equilibrium \cite{Sakharov:1967dj}. In the last couple of decades several attempts have been made to address the phenomena of BAU. One of the popular and successful theoretical process is baryogenesis via leptogenesis. This mechanism was first proposed by Fukugita and Yanagida \cite{Fukugita:1986hr}. According to this mechanism, the L-violating out of equilibrium decays of singlet neutrino creates an symmetry in the leptonic sector. This excess in lepton number can be converted into the observed baryon asymmetry through B+L violating sphaleron processes \cite{Trodden:2004mj,Kolb:1990vq}. In this regard, Inverse seesaw contains gauge singlet right-handed neutrinos and sterile fermions. As a result, the asymmetry generated by decay of one of the quasi-Dirac pairs formed by these particles can be converted into baryon asymmetry of the universe.

	The presence of dark matter (DM), an inevitable mystery of the universe, has been well-established through various observations in astrophysics and cosmology. Some of the strong observations in this regard are galaxy cluster observations by F. Zwicky \cite{Zwicky:1933gu}, galaxy rotation curves \cite{Rubin:1970zza}, recent observation of the Bullet clusters \cite{Clowe:2006eq} and cosmological data from the Planck collaboration \cite{Planck:2019nip}. All of these remarks suggest the existence of an unknown, non-luminous, non-baryonic dark matter which constitutes about 26\% of the energy density of the universe and is approximately five times more than luminous matter. Currently the amount of dark matter in the universe as found from the Planck data is \cite{Planck:2018vyg}
	\begin{equation}
		\Omega h^2=0.1199\pm0.0027
		\label{sec1eq2}
	\end{equation}
	This is called the relic density of dark matter. The properties that a candidate must have to qualify as dark matter candidate has been highlighted in \cite{Taoso:2007qk,Murayama:2007ek}. Unfortunately, the SM particles do not possess these required criteria and so none of them can be considered to be a viable dark matter candidate. Therefore , from the particle physics view point, one has to extend the SM particle content by incorporating new fields to find a suitable candidate that could produce the correct relic abundance.

	Another serious conundrum in particle physics is the nature of neutrinos \cite{Barenboim:2002hx,Czakon:1999cd,Bilenky:2012qi}: whether they are Dirac or Majorana particle. A possible solution to this question lies in the discovery of neutrinoless double beta decay (0$\nu\beta\beta$/NDBD). It is a SM forbidden second order lepton number violating interaction, which if discovered can shade light about the Majorana nature of the neutrinos. The decay channel of this reaction can be expressed as: $$(A,Z)\rightarrow (A,Z+2) + 2e^-$$
	
	Currently there are many ongoing experiments which actively look for this decay. Some of these popular experiments are KamLANDZen \cite{KamLAND-Zen:2016pfg}, GERDA \cite{GERDA:2020xhi}, CUROE etc. These experiments use Xenon-136 and Germanium-76 nuclei for possible detection of lepton number violating decays. The recent results published from these experiments predict the allowed lower bound for half-life ($T_{\frac{1}{2}}$) to be $\geq 10^{26}$ years. Moreover the effective neutrino mass is found to be $\leq$ 0.165 eV.

	Symmetry plays a very important role in model building in particle physics. Numerous work that has been done based on discrete flavor symmetry can be found in \cite{Borah:2017dmk,King:2009ap,King:2009db,Gautam:2020wsd}. In this type of symmetry a large number of flavons, along with their VEV alignments, are used in the models.The non-Abelian discrete symmetry groups like $A_4$, $S_4$, $Z_N$ etc. play a very crucial role in developing these models. In our work we have used modular symmetry in the framework of minimal Inverse seesaw [ISS(2,3)]. In this type of symmetry the Yukawa couplings are not free. Instead they are functions of a complex variable $\tau$, called the modulus \cite{Feruglio:2017spp}. As a result this helps in reducing the number of flavons that are used in the model. In addition to this, one does not have to deal with their VEV alignments. The modular group, $\Gamma(N)$ acts on the upper half of the complex plane and transforms $\tau$ in the following way: $$\tau\rightarrow \frac{a\tau+b}{c\tau+d}$$ 
	where $a,b,c,d$ are the elements of a 2$\times$2 matrix, such that $ad-bc=1$. The literatures \cite{Nomura:2019xsb,Behera:2021eut,Zhang:2021olk,Nomura:2020cog} highlight some of the recent works  done in inverse seesaw using modular symmetry. Some of the finite modular groups ( $N\leq5$) are isomorphic  to non-Abelian discrete symmetry groups ($A_4, S_4, A^\prime_5$ etc). In our work we have used $\Gamma(3)$ modular group. This group has three Yukawa modular forms of weight 2. One can obtain the higher weight modular forms by using these weight 2 modular functions. Moreover their values can be found with the help of their q-expansions \cite{Novichkov:2019sqv}. A brief description on $\Gamma(3)$ is given in Appendix \ref{appenA}. The authors in \cite{Mukherjee:2015axj,Verma:2021koo} have studied dark matter phenomena using discrete flavor symmetry. But in our work, along with dark matter, we have also studied leptogenesis and neutrinoless double beta decay by using modular symmetry.

	This paper is organised as follows: in section \ref{sec2} we discsuss the basic structure of the model, charge assignments of particles under various groups used in the work and then construct its lagrangian. In section \ref{sec3} and \ref{sec4} we give a brief description about neutrinoless double beta decay and leptogenesis in the framework of ISS(2,3). We have given a brief explanation about dark matter in section \ref{sec5}.	In section \ref{sec6} we provide the analysis and results of our work. We have shown the different plots and their descriptions in this part of the paper. Finally in section \ref{sec7} we conclude by giving an overview of our entire work.

		\section{The framework of inverse seesaw}
		
		Inverse seesaw is a popular BSM framework which is used in model building in neutrino physics. It is an extension of the Standard Model with three right-handed neutrinos and  gauge singlet sterile fermions. Unlike the conventional seesaw mechanism, the Majorana mass of right-handed neutrinos in Inverse seesaw can be reduced to TeV scale. As a result, apart from generating tiny neutrino masses, it also enhances the possibility of detecting these right-handed particles in the ongoing and near future experiments.These possibilities of inverse seesaw have been highlighted in \cite{Hirsch:2009mx,Gu:2010xc,Nomura:2018cfu,Awasthi:2013we,Malinsky:2009df}. For the basis $n_L=(\nu_{L,\alpha},N^c_{R,i},S_j)^T$, the Yukawa lagrangian  for neutrino in this framework can be written as:
		\begin{equation}
		\mathcal{L}=-\frac{1}{2}n^T_LCMn_L+h.c.
		\label{secIIeq1}
		\end{equation}
		
		where $C\equiv \gamma^2 \gamma^0$ is the charge conjugation matrix. The component $\nu_{L,\alpha}$ for $\alpha=e,\mu,\tau$ are left-handed Standard Model neutrinos. The complete mass matrix for neutral fermion arising from the lagrangian can be written as \cite{Abada:2014vea}:
		\begin{equation}
		\mathcal{M}= \begin{pmatrix}
		0&M^T_D&0\\M_D&0&M_{NS}\\0&M^T_{NS}&M_S
		\end{pmatrix}_{9\times9}
		\label{secIIeq2}
		\end{equation} 
		
		The elements of the matrix in eq. (\ref{secIIeq2}) represent the different mass matrices involved in ISS mechanism. The Dirac mass matrix ($M_D$) results from the interaction between left and right-handed components of neutrinos. $M_{NS}$ is the mixing matrix that occurs due to right-handed neutrinos and sterile singlet fermions interactions. The interaction between the singlet sterile fermions forms the Majorana mass term, $M_S$. The dimensions of the  matrices are related to the number of generations of the particles that have been considered in the model. Accordingly in ISS the dimensions of these matrices can be defined in the following way (\# represents the number of generations of particles.):
		\begin{equation}
			\begin{aligned}
		&	Dimension \hspace{1.5mm} of\hspace{1.5mm} M_D=(\#\nu_L\times \#N_R)\\
		&	Dimension \hspace{1.5mm} of\hspace{1.5mm} M_{NS}=(\#N_R\times \#S)\\
		&   Dimension \hspace{1.5mm} of\hspace{1.5mm} M_S=(\#S\times \#S)
			\end{aligned}
			\label{secIIeq3}
		\end{equation}

	From eq. (\ref{secIIeq3}) one can find that the three matrices, $M_D$, $M_{NS}$ and $M_S$, have the same dimension i.e. $3\times3$. The effective neutrino mass matrix for the active light neutrinos can be written as: \begin{equation}
	m_\nu=M^T_D(M^T_{NS})^{-1}M_SM^{-1}_{NS}M_D
	\label{secIIeq4}
	\end{equation}
		
		The light neutrino mass matrix in eq. (\ref{secIIeq4}) can be diagonalised with the help of an unitary matrix. The corresponding eigenvalues will be the mass of the active neutrinos. In order to produce sub-eV Standard Model neutrinos, $M_D$ must be in electroweak range, $M_{NS}$ in the TeV range and $M_S$ must be in the KeV range, respectively.

		\subsection{The Model}	\label{sec2}

	    In this work we have used minimal inverse seesaw [ISS(2,3)] framework. Compared to ISS, this mechanism of minimal inverse seesaw constitutes of two right-handed neutrinos ($N_1,N_2$) and three singlet sterile fermions ($S_1,S_2,S_3$). As a result, the order of $M_D$, $M_{NS}$ changes to $3\times2$ and $2\times3$; whereas for $M_S$ it remains unchanged. Along with these particles we have used a flavon ($\phi$) whose VEV alignments facilitate to get a diagonal charged lepton mass matrix. As such the role of this flavon is restricted only to the charged lepton sector without affecting the neutrino sector. We have extended the minimal inverse seesaw by a Higgs-type scalar triplet $\eta=(\eta_1,\eta_2,\eta_3)$ whose neutral component is our dark matter candidate \cite{Boucenna:2011tj,Hirsch:2010ru}. Proceeding ahead, we have used $A_4$ modular symmetry to construct the desired lagrangian of the model. This group contains three Yukawa modular forms ($y_1,y_2,y_3$) of weight 2 which have been considered as triplet under $A_4$ charge assignment.  The right-handed neutrinos ($N_1$ and $N_2$) in the model are taken to be singlets under $A_4$ and they transform as $1'$ and $1''$, respectively; while the sterile neutrinos ($S_i$) and lepton doublets (L) are considered as triplets. The modular weight of the right-handed neutrinos is taken as -2 whereas for the lepton doublets (L), sterile neutrinos and $\eta$ it is taken to be zero. In order to restrict certain interaction terms in the Lagrangian, we have used $Z_3$ symmetry group. Moreover we introduce two weighton fields, $\beta_1$ and $\beta_2$, which are associated with the right-handed neutrinos. These charge assignments of the particles under various groups in the model have been highlighted in table \ref{tab:1}. 
		
		\begin{table}[ht]
			\centering
			\begin{tabular}{|c| c c c c c c c c |}
				\hline
				Fields & L &  $N_1$ & $N_2$ & $S_i$ & H & $\phi$ & $\eta$ & Y \\
				\hline
				$A_4$ & 3 & $1^\prime$ & $1^{\prime\prime}$ & 3 & 1 & 1 & 3 & 3 \\
				
				$Z_3$ & $\omega^2$ & $\omega$ & $\omega$ & 1 & $\omega$ & $\omega$ & $\omega$ & $\omega^2$ \\
				$K_I$ & 0 & -2 & -2 & 0 & 0 & 2 & 0 & 2 \\
				\hline
			\end{tabular}
			\label{tab:1}
			\caption{Charge assignments of the particles under the various groups considered in the model.}
		\end{table}

		As we have considered a DM candidate in our model, it is necessary to introduce a discrete symmetry $Z_2$ in order to maintain  stability of the DM candidate.  The SM particles remain $Z_2$ even whereas the right handed neutrino and the newly added field eta are odd under this symmetry. The scalar potential of the Higgs sector can be found in \cite{Boucenna:2011tj}. After electroweak symmetry breaking, one of the $\eta$'s acquire VEV and their form can be written as:
		
		\begin{equation}
		\eta_1=\begin{pmatrix}
		\eta_1^+\\ \frac{v_\eta+h_1+iA_1}{\sqrt{2}} 
		\end{pmatrix},	\hspace{5mm}	\eta_2=\begin{pmatrix}
		\eta_2^+\\ \frac{h_2+iA_2}{\sqrt{2}}
		\end{pmatrix}, \hspace{5mm} 	\eta_3=\begin{pmatrix}
		\eta_3^+\\ \frac{h_3+iA_3}{\sqrt{2}}
		\end{pmatrix}
		\end{equation} 
		
		As mentioned in \cite{Hirsch:2010ru}, the VEV alignment of $\eta$ can be written as $\eta=v_\eta(1,0,0)$ and $\eta_2,\eta_3$ will be the dark matter candidates. Moreover we get a diagonal charge lepton mass matrix when the VEV of $\phi$ is taken as  $\phi=(v,0,0)$ \cite{Feruglio:2017spp}. Based on the above discussions and relevant charge assignmets, the Yukawa Lagrangian for the neutrino sector can be written as: 
		\begin{equation}
			\mathcal{L}= N_1(LY)_3\eta + N_2(LY)_3\eta + \beta_1N_1(SY)_{1''} + \beta_2N_2(SY)_{1'}+\mu_0(SS)_1
			\label{sec2eq1}
		\end{equation}

		The first two terms in the lagrangian denotes the interaction between left-handed and right-handed neutrinos. The next two terms represents the interaction between $N$'s and $S$'s while the last term denotes the interaction between the sterile fermions. As a result, from eq. (\ref{sec2eq1}) we can write down the corresponding neutrino mass matrices in the following way. The Dirac mass matrix, which is a matrix of order 3$\times$2, for the neutrinos can be written as: 
		
		\begin{equation}
			M_D= v_\eta\begin{pmatrix}
				-y_1 & -y_2\\2y_2 & -y_1\\-y_1 & 2y_3
			\end{pmatrix}
			\label{sec2eq2}
		\end{equation}
		Similarly, following the $A_4$ multiplication rules, the Majorana mass matrix for right-handed neutrino and sterile fermions, and the lepton number violating mass term for the sterile fermions can be written as:
		\begin{equation}
			M_{NS}=\begin{pmatrix}
				\beta_1y_3 & \beta_1y_2 & \beta_1y_1\\ \beta_2y_2 & \beta_2y_1 & \beta_2y_3
			\end{pmatrix}, \hspace{7mm} 	M_S=\mu_0 \begin{pmatrix}
			1 & 0 & 0\\ 0 & 0 & 1\\ 0 & 1 & 0
			\end{pmatrix}
			\label{sec2eq3}
		\end{equation}

		Following the methods as mentioned in the literature, the full $8\times 8$ neutrino mass matrix for ISS(2,3) can be written as:
		\begin{equation}
			\mathcal{M}= \begin{pmatrix}
				0&M^T_D&0\\M_D&0&M_{NS}\\0&M^T_{NS}&M_S
			\end{pmatrix}_{8\times8}
			\label{sec2eq5}
		\end{equation}

		This $8\times8$ matrix $M$ in eq (\ref{sec2eq5}) can be diagonalised with the help of an Unitary matrix, $\mathcal{U}$ as
		\begin{equation}
			\mathcal{U^T}M\mathcal{U}=M_{diag}=diag(m_1,m_2,m_3,....,m_8)
			\label{sec2eq6}
		\end{equation}
		where $m_i$'s in the above equation are masses of the particles of the model. Since the matrix in eq (\ref{sec2eq3}) is a rectangular matrix, so we cannot find the light neutrino mass matrix by the conventional approach \cite{Abada:2017ieq}. Because of this reason, the formula to find the effective light neutrino mass matrix in eq, (\ref{secIIeq4}) changes and takes the form \cite{Abada:2014zra}:
		\begin{equation}
			m_\nu=M_D.d.M^T_D
			\label{sec2eq7}
		\end{equation}
		
		In the above equation, $d$ is a $2\times 2$ matrix which can be derived from the $5\times 5$ heavy neutrino mass matrix $M_H$. The form of $d$ can be obtained in the following way:
		\begin{equation}
			M_H^{-1}=\begin{pmatrix}
				0 & M_{NS} \\ M_{NS}^T & M_S
			\end{pmatrix}^{-1}=\begin{pmatrix}
				d_{2\times 2} & .....\\...... & .....
			\end{pmatrix}
			\label{sec2eq8}
		\end{equation}
		
		Thus we have constructed a model using $A_4$ modular symmetry in the framework of ISS(2,3). We can diagonalise eq. (\ref{sec2eq7}) to obtain the masses of light neutrinos which will facilitate to study the related neutrino phenomenology.

		%%%%%%%%%%%%%%%% new section %%%%%%%%%%%%%%%%%%%%%%%%%%%%%%%%%%%%%%%%%%%%%%%%%%%%%%%%%%
		
		\section{Leptogenesis}	\label{sec3}
		
		In order to validate our model considering the cosmological constraints, we try to generate the observed baryon asymmetry of the universe through leptogenesis. There are five heavy neutrinos in ISS(2,3). As mentioned above, two of them are right handed neutrinos ($N_1,N_2$) and the other three are gauge singlet neutral sterile fermions ($S_i$). Four of these heavy particles form two pairs, called quasi-Dirac pairs, and one of them gets decoupled. Interestingly, the mass splitting between these pairs is comparable to their decay width. The out-of-equilibrium decay of the lightest pair to any lepton flavor creates an asymmetry in the leptonic sector. This asymmetry created by decay of the lightest heavy neutrinos can be converted into baryon asymmetry through sphaleron processes \cite{Awasthi:2013we,Lindner:2014oea,Lucente:2018uaj}. On the other hand, asymmetry generated by decay of the heavier pair is washed out by lepton number violating scatterings of the lightest pair, thereby, it does not contribute to the
		asymmetry produced.
		
		\subsection{Computation of CP asymmetry}	
		To calculate the CP-asymmetry we need the mass matrix for the heavy neutrinos. This matrix can be written as:
		\begin{equation}
			M_H=\begin{pmatrix}
				0 & M_{NS} \\ M_{NS}^T & M_S
			\end{pmatrix}
			\label{sec3eq1}
		\end{equation}
		
		Now on diagonalising the above matrix, one can get masses of the five heavy neutrinos. Also the mass splitting between the degenerate pairs is proportional to $M_S$.
		\begin{equation}
			M_{diag}=V^TM_HV=diag(m_1,m_2,m_3,m_4,m_5)
			\label{sec3eq2}
		\end{equation}
		
		To diagonalise this $5\times 5$ matrix analytically is a challenging and formidable task. So we opt for numerical diagonalisation so as to simplify our analysis. Again for the calculation of CP-asymmetry, a particular basis is preferred in which the matrix $M_H$ becomes diagonal. In this basis the Lagrangian takes the form:
		\begin{equation}
			\mathcal{L_S}= h_{i\alpha}N_i\eta L_\alpha + M_iN_i^TC^{-1}N_i+h.c.
			\label{sec3eq3}
		\end{equation}
		
		where $h_{i\alpha}$ corresponds to the couplings in diagonal mass basis. These are related to the couplings in the flavor basis represented by the following relations \cite{Chakraborty:2021azg,Agashe:2018cuf}:
		\begin{equation}
			\begin{aligned}
				h_{1\alpha}&=V_{11}^*\hspace{1mm}y_{1\alpha}+V_{12}^*\hspace{1mm}y_{2\alpha}\\
				h_{2\alpha}&=V_{21}^*\hspace{1mm}y_{1\alpha}+V_{22}^*\hspace{1mm}y_{2\alpha}\\
				h_{3\alpha}&=V_{13}^*\hspace{1mm}y_{1\alpha}+V_{23}^*\hspace{1mm}y_{2\alpha}\\
				h_{4\alpha}&=V_{14}^*\hspace{1mm}y_{1\alpha}+V_{24}^*\hspace{1mm} y_{2\alpha}
			\end{aligned}
			\label{sec3eq4}
		\end{equation}

		For the decay $N_i\rightarrow l_\alpha \phi\hspace{2mm} (\bar{l_\alpha}\phi)$, the formula for calculating the CP asymmetry $\epsilon_i$ by summing over the SM flavor $\alpha$ is given by \cite{Covi:1996wh}
		\begin{equation}
			\epsilon_i=\frac{\sum_\alpha[\Gamma(\tilde{\psi}\rightarrow l_\alpha\phi)-\Gamma(\tilde{\psi}\rightarrow \bar{l_\alpha}\phi^\dagger)]}{\sum_\alpha[\Gamma(\tilde{\psi}\rightarrow l_\alpha\phi)+\Gamma(\tilde{\psi}\rightarrow \bar{l_\alpha}\phi^\dagger)]}=\frac{1}{8\pi}\sum_{i\neq j}\frac{Im[(hh^\dagger)^2_{ij}]}{(hh^\dagger)_{ii}}f_{ij}
			\label{sec3eq5}
		\end{equation}
		For the case of resonant leptogenesis, the self energy correction, $f_{ij}=\frac{(M_i^2-M_j^2)M_iM_j}{(M_i^2-M_j^2)^2+(M_i\Gamma_i+M_j\Gamma_j)^2}$. $M_i$ and $M_j$ are the real and positive eigenvalues of the heavy neutrino mass matrix. $\Gamma_i$ is the decay width of one of the quasi-Dirac pair which is expressed as $\Gamma_i=\frac{M_i}{8\pi}(hh)^\dagger_{ii}$. Thus the explicit form of CP parameter for the decay of a quasi-Dirac pair, say $(N_1,S_1)$, can be expressed as:
		\begin{equation}
			\begin{aligned}
				\epsilon_1 & =\frac{1}{8\pi (hh^\dagger)_{11}}\textnormal{Im}[(hh^\dagger)^2_{12}f_{12}+(hh^\dagger)^2_{13}f_{13}+(hh^\dagger)^2_{14}f_{14}]\\
				\epsilon_2 & =\frac{1}{8\pi (hh^\dagger)_{22}}\textnormal{Im}[(hh^\dagger)^2_{21}f_{21}+(hh^\dagger)^2_{23}f_{23}+(hh^\dagger)^2_{24}f_{24}]
			\end{aligned}
			\label{sec3eq6}
		\end{equation}
		
		As already mentioned, the asymmetry produced by decay of the heavier pair is washed
		out. Thus the wash out parameter for such decays, in terms of the Hubble parameter H, is written as:
		\begin{equation}
			K_i=\frac{\Gamma_i}{H}=\frac{M_i}{8\pi}(hh)^\dagger_{ii}\times\frac{M_{pl}}{1.66\sqrt{g*}M_i^2}
			\label{sec3eq7}
		\end{equation}
		
		In the above equation, $M_{pl}$ is the Planck mass and $g*$ denotes the effective number of relativistic degrees of freedom. The final expression for BAU can be written as:
		\begin{equation}
			Y_{B}=10^{-2}\sum \kappa_i\epsilon_i
			\label{sec3eq8}
		\end{equation}
		where $\kappa_i$ is the dilution factor responsible for washout out of the asymmetry associated with the heavy pair. $\epsilon_i$ is the CP asymmetry generated in the leptonic sector. The expressions for $\kappa_i$ depends on the values of washout factor in eq (\ref{sec3eq7}). These relations can be summarised as \cite{Pilaftsis:2005rv}:
		\begin{equation}
			\begin{aligned}
				-\kappa &\approx \sqrt{0.1K}\hspace{2mm}exp[\frac{-4}{3(0.1K)^{0.25}}], \hspace{6mm} \textnormal{for} K\geq10^6\\
				&\approx \frac{0.3}{K(\textnormal{ln}K)^{0.6}},\hspace{2.5cm} \textnormal{for} \hspace{2mm} 10\leq K\leq 10^6\\
				& \approx \frac{1}{2\sqrt{K^2+9}}, \hspace{2.5cm} \textnormal{for} \hspace{2mm} 0\leq K\leq 10
			\end{aligned}
			\label{sec3eq9}
		\end{equation}
		Thus, we use eq (\ref{sec3eq8}) to calculate the value of BAU in the framework of ISS(2,3). In our work we have computed the values of this asymmetry for both the hierarchies. We have highlighted our findings in the numerical analysis section of the paper. We find that for both the hierarchies there are a large number of points that satisfy the Planck value of BAU.

		\section{Neutrinoless double beta decay}\label{sec4}
		
		Neutrinoless double beta decay $(0\beta\beta\nu/\textnormal{NDBD})$ is an important phenomena in particle physics. It is a lepton number violating decay which if discovered can give an answer to many of the open questions in particle physics. In our work, we have studied the effective electron neutrino Majorana mass which characterizes $0\beta\beta\nu$. As mentioned in \cite{Blennow:2010th}, the experiments like KamLAND-Zen\cite{KamLAND-Zen:2016pfg}, GERDA\cite{GERDA:2020xhi}, CUORE\cite{CUORE:2019yfd} etc. provide a stringent bound on $m_{ee}$. In absence of any sterile neutrino, this effective mass is expressed as given below:
		\begin{equation}
			m_{ee}=\big|\sum^3_iU_{ei}^2m_i\big|
			\label{sec4eq1}
		\end{equation}
		The presence of sterile neutrinos in ISS(2,3) changes the expression for effective mass in eq (\ref{sec4eq1}). As a result, the modified expression become \cite{Gogoi:2022jwf,Abada:2018qok}:
		\begin{equation}
			m_{ee}=\big|\sum^3_{i=1}U_{ei}^2m_i\big|+ \big|\sum_{j=1}^5U_{ej}^2\frac{M_j}{k^2+M_j^2}\big|\big<k\big>\big|^2\big|
			\label{sec4eq2}
		\end{equation}
		where $U_{ej}$ represents the coupling of the heavy neutrinos to the electron neutrino. $M_j$ is the mass of the heavy neutrinos. The parameter $k$ is known as the virtuality momentum and its value is  $\big|\big<k\big>\big|$$\approx$ 190 MeV.

		\section{Dark Matter}\label{sec5}
		
		At the beginning of the universe the particles present in the thermal pool were in thermal equilibrium with each other. This implies that the rate at which lighter particles combined to form heavy particles and vice-versa was the same. During the course of evolution, the conditions that were required to maintain this equilibrium state were disturbed. As a result, after a certain temperature the density of some particle species became too low. Once this is achieved, the abundance of those particle remains the same and their density becomes constant. This phase of the particle species is called freeze-out and the density hereafter is referred to as relic density. For a particle $\chi$, which was in thermal equilibrium, the relic density can be obtained from the Boltzman equations \cite{Griest:1990kh,Gondolo:1990dk}:
		\begin{equation}
		\frac{dn_\chi}{dt}+3\mathcal{H}n_\chi=-\textless\sigma v\textgreater\left( n_\chi^2-\left( n_\chi^{eqb}\right)^2\right) 
		\label{sec5eq1}
		\end{equation}

		In the above equation $n_\chi$ is the number density of the dark matter particle whereas $n_\chi^{eqb}$ is the density of $\chi$ when it was in equilibrium with the thermal bath. Here $\mathcal{H}$ represents the Hubble constant and $\textless\sigma v\textgreater$ is the thermally averaged annihilation cross-section of the dark matter candidate. For the interactions in eq. (\ref{sec2eq1}), the cross-section formula can be writtern as \cite{Bai:2013iqa}:
		\begin{equation}
			\textless\sigma v\textgreater = \frac{v^2 y^4 m_\chi^2}{48\pi(m_\chi^2+m_\psi^2)^2} 
		\label{sec5eq2}
		\end{equation}
		
		where the parameters $m_\chi, m_\psi \hspace{2mm}\textnormal{and}\hspace{2mm} y$ represent the mass of the relic particle, mass of the Majorana fermion and the interaction between the dark matter and fermions in the model. $v$ in the above expression represents the relative velocity of the relic particles whose value at the time of freeze-out is taken to be $0.3c$. The solution of eq. (\ref{sec5eq1}) in terms of reduced Hubble constant ($h$), as found in \cite{Jungman:1995df}, can be written as:
		\begin{equation}
			\Omega_\chi h^2=\frac{3\times 10^{-27} \textnormal{cm}^3 \textnormal{s}^{-1}}{\textless \sigma v\textgreater}
			\label{sec5eq3}
		\end{equation}
		
		The above expression of $\Omega_\chi h^2$ gives the relic density of dark matter particle. It is found that the self annihilation between dark matter and next to lightest neutral component of $\eta$ contributes to the annihilation cross-section. For low mass region i.e. $m_{DM}\textless M_W$, the cross-section for the self annihilation of either $\eta_2$ or $\eta_3$ into Standard Model particles via the Higgs boson can be written as \cite{Bell:2013wua}:
		\begin{equation}
		\sigma_{xx}=\frac{|Y_f|^2|\lambda_x|^2}{16\pi s}\frac{(s-4m_f^2)^{\frac{3}{2}}}{\sqrt{s-4m_x^2}((s-m_h^2)^2+m_h^2\Gamma_h^2)}
		\label{sec5eq4}
		\end{equation}
		
		where $x$ is the dark matter particle ($\eta_2,\eta_3$) and $\lambda_x$ is the coupling of $x$ with the SM Higgs boson $h$. In eq. (\ref{sec5eq4}) $Y_f$ represents the Yukawa couplings of the fermions. $\Gamma_h$ represents the decay width of the SM Higgs and $m_h$ is equal to 125 GeV. And $s$  in the expression represents the thermally averaged center of mass squared energy and is given by $s=4m_x^2+m_x^2v^2$, $m_x$ is mass of the relic. In our work the neutral component of the scalar triplet $\eta$ is the dark matter candidate and its mass should be less than the mass of $W$ boson, $M_W$.  The results that we have found in this work and their respective analysis are shown in the next section.

		\section{Numerical Analysis}\label{sec6}
		
		In this section we discuss the processes that we followed to arrive at the results of our work. In our calculations we have used the latest nu-fit data for the oscillation parameters. This set of data has been shown in the table below:
		
		\begin{table}[ht]
			\begin{tabular}{|c | c | c|}
				\hline
				Oscillation parameters & Normal Ordering & Inverted Ordering \\
				\hline
				$\textnormal{sin}^2\theta_{12}$ & [0.269,0.343] & [0.269,0.343]\\
				\hline
				$\textnormal{sin}^2\theta_{23}$ & [0.407,0.618] & [0.411,0.621]\\
				\hline
				$\textnormal{sin}^2\theta_{13}$ & [0.02034,0.02430] & [0.020530,0.02436] \\
				\hline
				$\Delta m^2_{21}/10^{-5} \textnormal{eV}^2$ & [6.28,8.04] & [6.82,8.04] \\
				\hline
				$\Delta m^2_{31}/10^{-3} \textnormal{eV}^2$ & [2.431,2.598] & [2.412,2.583]\\
				\hline
			\end{tabular}
		\caption{The latest $3 \sigma$ nu-fit values of oscillation parameters \cite{Esteban:2020cvm}.}
		\end{table}

		In order to find the model parameters, we diagonalise the light neutrino mass matrix using the standard relation $m_\nu=U^T \textnormal{diag}(m_1,m_2,m_3)U$, where $U$ is the PMNS matrix. In normal hierarchy the diagonal matrix can be written as $\textnormal{diag}(0,\sqrt{m_1^2+\Delta m^2_{solar}},\sqrt{m_1^2+\Delta m^2_{atm}})$, whereas in the inverted ordering it takes the form $\textnormal{diag}(\sqrt{m_3^2+\Delta m^2_{atm}}),\sqrt{\Delta m_{atm}^2+\Delta m^2_{solar}},0)$ \cite{Nath:2016mts}. Through this process we get the masses of light active neutrinos. Moreover, the solar and atmospheric mass squared differences, along with the sum of neutrino masses, provide additional constraints on the model. In order to evaluate the values of the mixing angles from the model we use the following relations \cite{Nomura:2019xsb}:
		
		\begin{align}
			\textnormal{sin}^2\theta_{13}=|U_{e3}|^2,   \hspace{6mm} \textnormal{sin}^2\theta_{23}=\frac{|U_{\mu 3}|^2}{1-|U_{e3}|^2}, \hspace{6mm} \textnormal{sin}^2\theta_{12}=\frac{|U_{e 2}|^2}{1-|U_{e3}|^2},
		\end{align}
		
		The real and imaginary parts of the complex modulus, $\tau$ is found to lie within the fundamental domain \cite{Feruglio:2017spp}. Here in this work the ranges for the real and imaginary parts are found to be: Re($\tau$)$\rightarrow$[-0.9,\hspace{1mm}0.9] and Im($\tau$)$\rightarrow$[0.3,\hspace{1mm}6]. With these values of $\tau$ we find the Yukawa modular forms for both normal and inverted hierarchies. These values of the couplings have been highlighted in table \ref{tabyukawavalues}. For further calculations we have taken $\Lambda$ in the range [10,20] KeV and $v_\eta$ is considered in between ($30-50$) GeV. In a similar way, we have taken the values of $\beta_1$ and $\beta_2$ in the ranges [$10^5,10^7$] GeV and [$10^2, 10^3$] GeV, respectively, so as to obtain the values of different parameters in the allowed ranges. In the next part of this section we will show the different plots and then discuss the respective results that are obtained in this work.
		
		\begin{table}[ht]
			\centering
			\begin{tabular}{|c | c | c|}
				\hline
				Yukawa couplings & Normal Hierarchy & Inverted Hierarchy \\
				\hline
				|$y_1$| &  0.93 - 2   & 0.94 - 2.5    \\
				\hline
				|$y_2$| &   0.5 - 1   &  0.01 - 0.6   \\
				\hline
				|$y_3$| &   0.1 - 0.7   &  0.01 - 0.5   \\
				\hline
			\end{tabular}
		\caption{Values of the Yukawa modular forms.}
		\label{tabyukawavalues}
		\end{table}
		
\begin{figure}[H]
			\centering
			\includegraphics[scale=0.35]{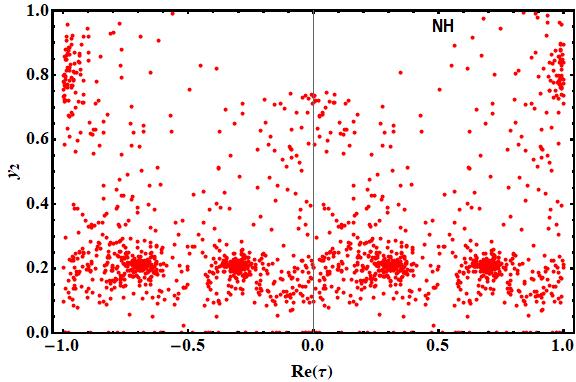}
			\hspace{5mm}
			\includegraphics[scale=0.35]{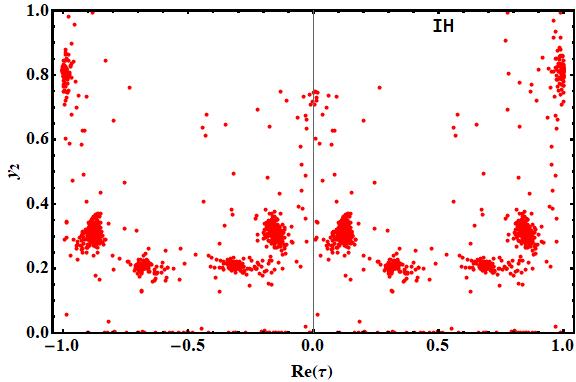}
			\caption{Corelation between Re($\tau$) and Yukawa modular form $y_2$ for both normal and inverted hierarchies.}
			\label{fig1}
		\end{figure}

\begin{figure}[H]
		\centering
		\includegraphics[scale=0.35]{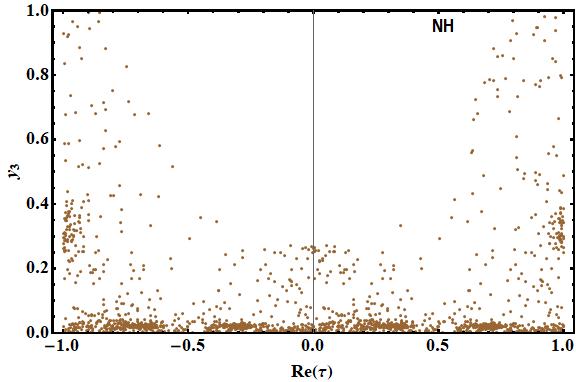}
		\hspace{5mm}
		\includegraphics[scale=0.35]{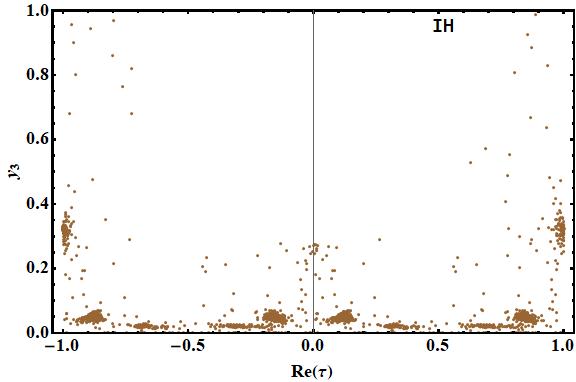}
		\caption{The above figure shows the relation between Re($\tau$) and $y_3$ for both the hierarchies.}
		\label{fig2}
	\end{figure}

\begin{figure}[H]
			\centering
			\includegraphics[scale=0.35]{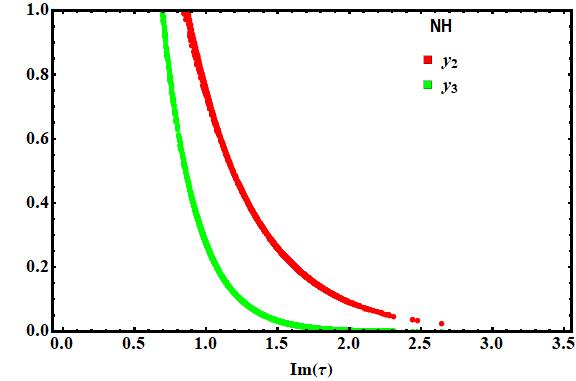}
			\hspace{5mm}
			\includegraphics[scale=0.35]{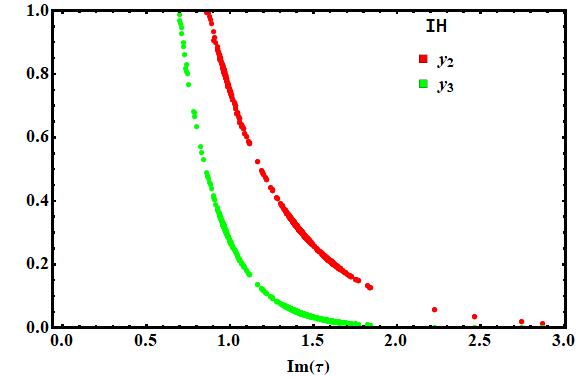}
			\caption{Corelation between Im($\tau$) and the Yukawa modular forms ($y_2$ and $y_3$).}
			\label{fig3}
		\end{figure}

		In fig.(\ref{fig1}) and (\ref{fig2}), we show the relation between the Yukawa modular forms and real part of the modulus, $\tau$.  In case of normal hierarchy, the values of Yukawa coupling ($y_2$) is found to lie mainly around the region (0.1-0.3) and for $y_3$ the parameter space is about 0.1. Some patterns of clusters can be found between the parameters in these zones. These patterns are more prominent in the case of inverted hierarchy. In the case of inverted hierarchy, however, the pattern is almost similar to NH. Fig. (\ref{fig3}) shows the relation between the imaginary part of $\tau$ and the Yukawa modular forms. It is evident from the graphs that most of the values of the couplings lie in between (0.9-2.5) of Im($\tau$). 
		
\begin{figure}[H]
			\centering
			\includegraphics[scale=0.35]{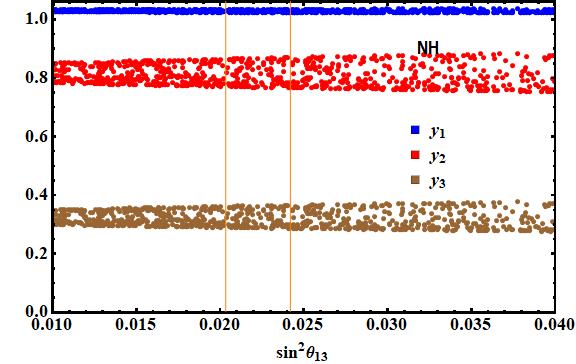}
			\hspace{5mm}
			\includegraphics[scale=0.35]{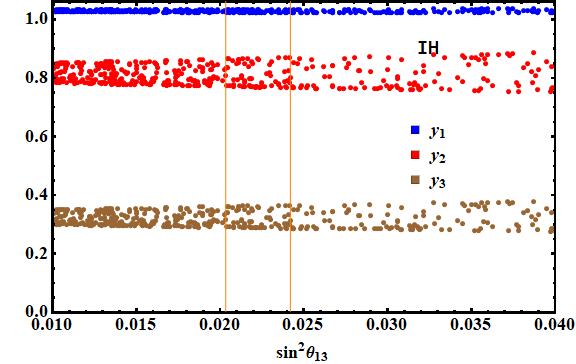}
			\caption{Corelation between sin$^2\theta_{13}$ and yukawa couplings ($y_1,y_2,y_3$).}
			\label{fig4}
		\end{figure}

\begin{figure}[H]
			\centering
			\includegraphics[scale=0.35]{sin12yukawa}
			\hspace{5mm}
			\includegraphics[scale=0.35]{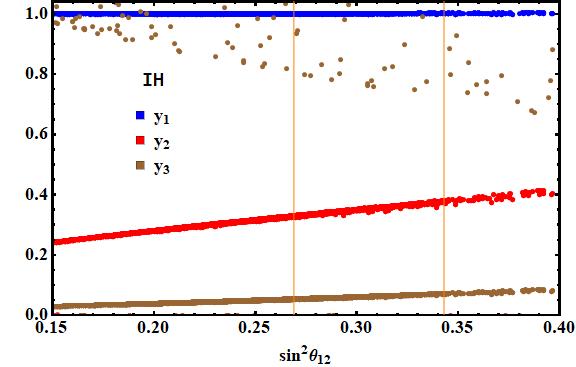}
			\caption{Corelation between sin$^2\theta_{12}$ and yukawa couplings ($y_1,y_2,y_3$).}
			\label{fig5}
		\end{figure}

\begin{figure}[H]
			\centering
			\includegraphics[scale=0.35]{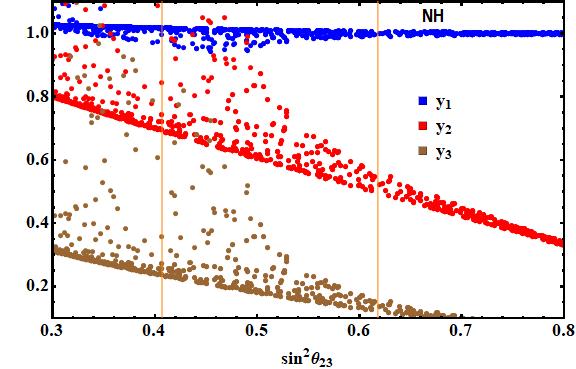}
			\hspace{5mm}
			\includegraphics[scale=0.35]{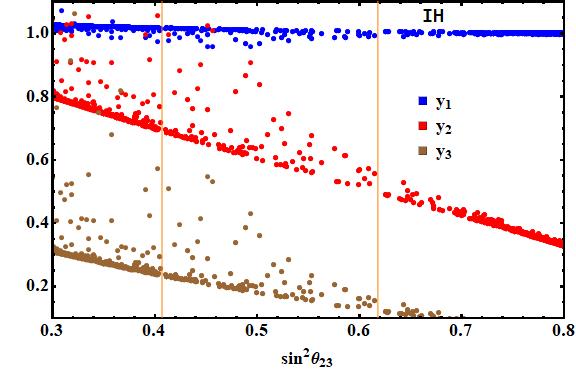}
			\caption{Corelation between sin$^2\theta_{23}$ and yukawa couplings ($y_1,y_2,y_3$).}
			\label{fig6}
		\end{figure}

		Next we discuss the relation between the mixing angles and modular forms used in this work. Fig. (\ref{fig4}), (\ref{fig5}), and (\ref{fig6}) highlight these relations for both the normal and inverted hierarchies. As evident from these graphs, there are sufficient parameter spaces that lie within the allowed ranges. In fig. (\ref{fig4}) we can see the relation with respect to $\textnormal{sin}^2\theta_{13}$.
	
	 There are three prominent regions of the Yukawa couplings that give the mixing angle in the allowed range. These ranges can be identified as:  $0.99\leq |y_1| \leq 1.2$, $0.78\leq |y_2| \leq 0.87$ and $0.29\leq |y_3|\leq 0.35$. Similarly we can find the ranges for the other graphs also. From fig. (\ref{fig5}) we can see that the allowed range of sin$^2\theta_{12}$ is prominent for the values of Yukawa couplings in the range: $0.99\leq |y_1| \leq 1.2$, $0.25\leq |y_2| \leq 0.35$ and $0.05\leq |y_3|\leq 0.1$. A few values of $y_3$ are also found in the upper part of the graphs. In case of atmospheric mixing angle (sin$^2\theta_{23}$), the ranges of the Yukawa couplings from fig. (\ref{fig6}) can be found as: $0.98\leq |y_1| \leq 1.2$, $0.5\leq |y_2| \leq 0.7$ and $0.15\leq |y_3|\leq 0.25$. Moreover there are few values of $y_2$ and $y_3$ which are randomly spread in the upper and middle portion of the graphs. This shows that there are certain regions of Yukawa modular forms which are more favorable at producing the mixing angles. In the following figures, the vertical lines represent the 3$\sigma$ bounds of the mixing angles for both the normal and inverted hierarchies. The points in the region within these lines represent the allowed ranges of the parameters. 
		
\begin{figure}[H]
			\centering
			\includegraphics[scale=0.35]{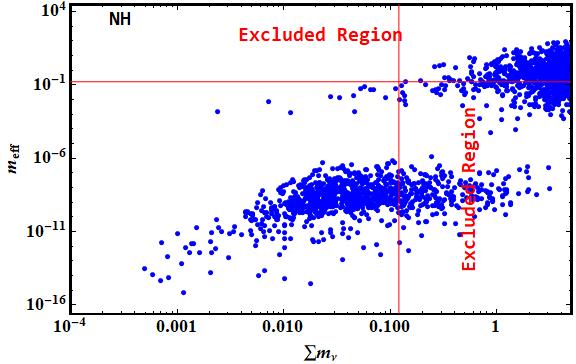}
			\hspace{5mm}
			\includegraphics[scale=0.35]{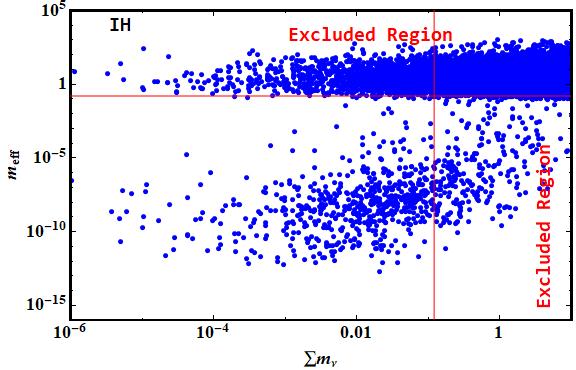}
			\caption{The above figures illustrate the corelation between $\sum m_\nu$ and $m_{eff}$. The horizontal and vertical lines represent the upper bounds of the parameters. }
			\label{fig7}
		\end{figure}

		Fig. (\ref{fig7}) shows the relation between sum of neutrino mass ($\sum m_\nu$) and effecive mass ($m_{eff}$). The recent cosmological findings from KamLANDZen provide the upper bound of $\sum m_\nu$ $\leq$ 0.12 eV, whereas the allowed range for the effective neutrino mass for NDBD is found to be $\leq$ 0.165 eV. The horizontal line in the figure represents the upper bound of effective electron neutrino mass, whereas the vertical line represents the upper bound for sum of neutrino mass. It is quite evident from both the graphs that there are sufficient parameter space within the allowed range for both the normal and inverted hierarchies.

\begin{figure}[H]
			\centering
			\includegraphics[scale=0.35]{baum1}	
			\hspace{5mm}
			\includegraphics[scale=0.35]{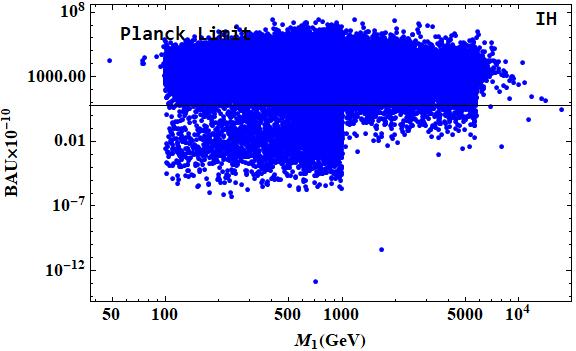}
			\caption{The figures above show the corelation between BAU and $M_1$ for both the hierarchies. In these figures, the horizontal lines represent the Planck value of BAU. }
			\label{fig8}
		\end{figure}

\begin{figure}[H]
			\centering
			\includegraphics[scale=0.35]{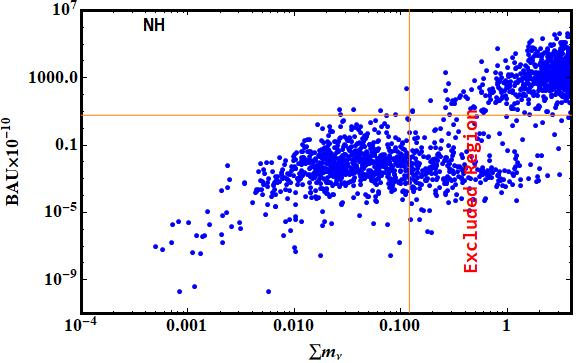}
			\hspace{5mm}
			\includegraphics[scale=0.35]{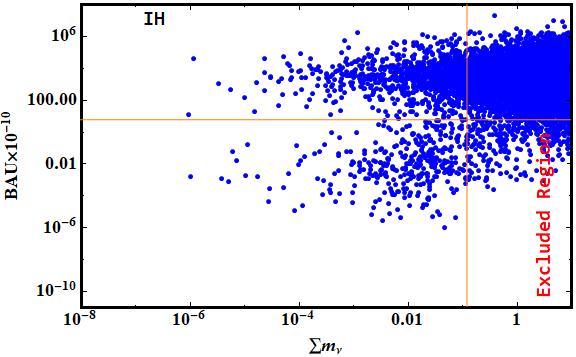}
			\caption{These figures show the corelation between BAU and sum of neutrino masees $\sum m_{\nu}$. The horizontal and vertical lines represent the Planck value of BAU and upper bound of $\sum m_\nu$, respectively.}
			\label{fig9}
		\end{figure}

		In fig. (\ref{fig8}) we show the relation between BAU and lightest right-handed neutrino mass, $M_1$. We performed the calculations for both the hierarchies. From these graphs it is clear that for both the type of hierarchies there are sufficient values of BAU which satisfy the Planck limit. This value of BAU for normal ordering is mainly concentrated in between (100-1000) GeV of $M_1$. Beyond this mass range the values of asymmetry is very less. But for inverted ordering this mass range is found to lie in between (100-5000) GeV. From this discussion it is clear that the values of baryon asymmetry of the universe can be obtained for both the hierarchies from this model. Also in fig. (\ref{fig9}), we show the relation between BAU and sum of neutrino masses.

\begin{figure}[H]
		\centering
		\includegraphics[scale=0.35]{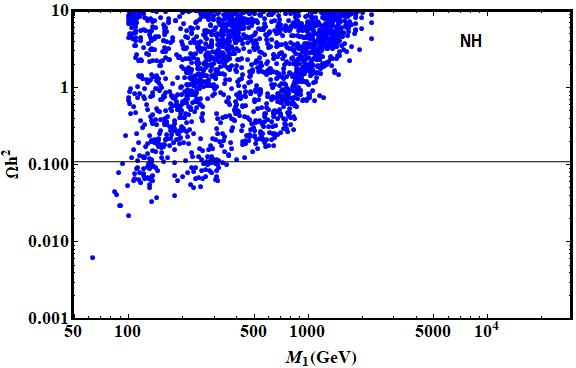}
		\hspace{5mm}
		\includegraphics[scale=0.35]{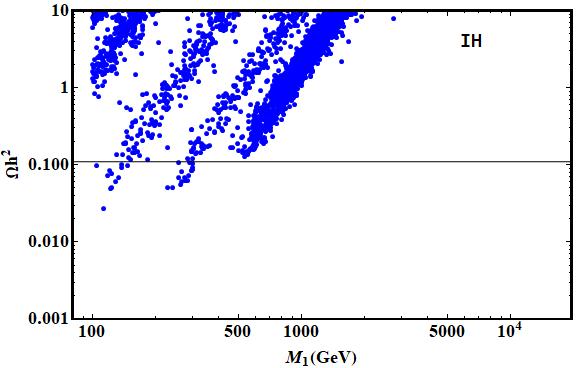}
		\caption{The above figure shows the variation between DM relic density and lightest right-handed neutrino $M_1$. The horizontal line represents the current dark matter abundance in the universe.}
		\label{fig10}
		\end{figure}

\begin{figure}[H]
		\centering
		\includegraphics[scale=0.35]{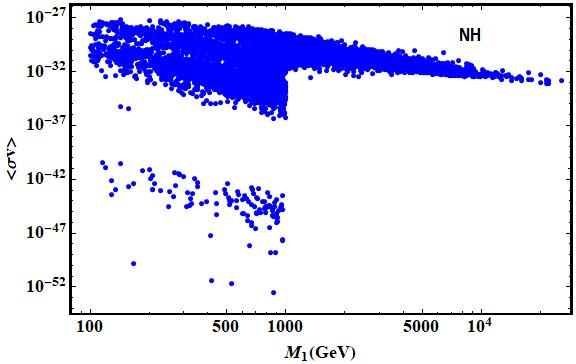}
		\hspace{5mm}
		\includegraphics[scale=0.35]{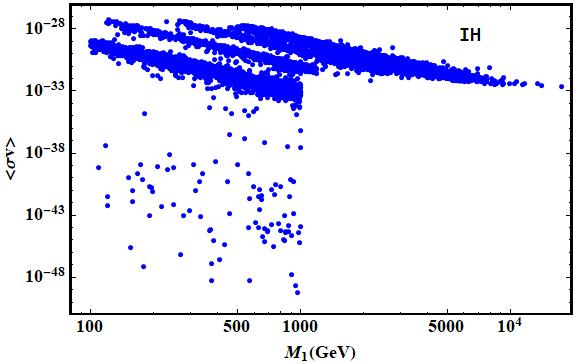}
		\caption{The above figure shows the corelation between cross-section $\textless\sigma v\textgreater$ and $M_1$.}
		\label{fig11}
	\end{figure}

	Finally in fig. (\ref{fig10}) and (\ref{fig11}) we show the variation of dark matter relic density and thermally averaged annihilation cross-section with respect to lightest heavy neutrino $M_1$. It can be seen that for $M_1$ in between (100-450) GeV, the values of relic density is more prominent for both the hierarchies. However, in case of inverted hierarchy, there are certain areas of mass that produce the observed relic density. As for the thermally averaged cross-section, we see that for almost the entire mass range of the $M_1$, the values of the scattering cross section is consistent with the experimental upper bounds \cite{Fermi-LAT:2010qeq}.

\section{Conclusion}\label{sec7}
         
         In this work, we have extended the minimal inverse seesaw by adding an extra Higgs-type scalar triplet ($\eta$) to its particle content. As mentioned in the earlier sections, this is a very popular mechanism in model building bescause of its ability to reduce the mass scale of right-handed neutrinos to TeV. We have used $A_4$ modular symmetry and $Z_3$ symmetry groups in our work. In $A_4$ modular group there are three Yukawa modular forms which are functions of the complex modulus $\tau$. These modular forms play a vital role in our calculations. In this paper, we have studied neutrino masses and mixings, neutrinoless double beta decay, baryogenesis via leptogenesis and dark matter. We have considered the results and bounds related to these phenomena which are published by different ongoing experiments and observations. We find that the values of the parameters calculated from the model are within the allowed ranges. Moreover, we have performed these calculations for both the normal and inverted hierarchies. In case of the Yukawa couplings, most of the values are mainly found to lie in the region (0.01 - 2). Also there is a sufficient parameter space within the allowed range for sum of neutrino masses and effective electron neutrino mass for NDBD.

         Furthermore, on studying BAU in our model, we have obtained a satisfying parameter space corresponding to $M_1$ which abide by the Planck limit for both NO/IO. Again, we have extended our investigation to dark matter sector as well by calculating the relic abundance and scattering cross section of the DM candidate. From plots Fig. 10 and 11, we have obtained a certain parameter space for DM mass which generates the observed relic abundance. Also, we have sufficient points which showcase very small scattering cross section. Thus, from the study of various phenomena that we have carried out in this work, we can have a conclusive idea that this model is a viable one.

\revappendix

\section{Modular Symmetry}	\label{appenA}

In recent times modular symmetry has become quite popular in the study of neutrino phenomenology. The modular group $\Gamma(N)$ ($N=1,2,3..$) can be defined in the following way:
\begin{equation}
\Gamma(N)=\left\{\begin{pmatrix} a & b\\c & d \end{pmatrix} \in SL(2,Z) , \begin{pmatrix} a & b\\c & d \end{pmatrix}=\begin{pmatrix} 1 & 0\\0 & 1 \end{pmatrix} (\textnormal{mod N})\right\}
\label{appeeqn1}
\end{equation} such that $ad-bc=1.$ These groups act on the upper half of the complex plane, (Im$(\tau)\textgreater0$) and transforms the complex variable $\tau$ linearly as:
$$\tau \rightarrow \frac{a\tau + b}{c\tau+d}$$

The matrix form of the two generators of modular symmetry are:

\begin{align}
S=\begin{pmatrix} 0 & 1\\-1 & 0\end{pmatrix} ,  & \hspace{2cm}  T=\begin{pmatrix} 1 & 1\\0 & 1\end{pmatrix}.
\label{appeneq2}
\end{align} These operators act on $\tau$ and transform them in the following ways:

\begin{align}
S&\xrightarrow{\tau} -\frac{1}{\tau}, & T&\xrightarrow{\tau} 1+\tau.
\label{appeneeq3}
\end{align}  
 The finite modular groups $(N\leq5)$ and non-abelian discrete groups are isomorphic to each other. As a result $\Gamma_2\approx S_3$, $\Gamma_3\approx A_4$,$\Gamma_4\approx S_4$,$\Gamma_5\approx A'_5.$ For a group of level $N$, the number of modular forms varies with respect to their weights. The table below shows how to find out the number of modular forms that a particular group with a particular level can have.

 \begin{table}[H]
 	\centering
 	\begin{tabular}{|c|c|c|}
 		\hline
 		$N$ &  No. of modular forms& $\Gamma(N)$\\
 		\hline
 		2 & $k+1$ & $S_3$\\
 		\hline
 		3 & $2k+1$ & $A_4$\\
 		\hline
 		4 & $4k+1$ & $S_4$\\
 		\hline
 		5 & $10k+1$ & $A_5$\\
 		\hline
 		6 & $12k$ &\\
 		\hline
 		7& $28k-2$ &\\
 		\hline
 		
 	\end{tabular}
 	\caption{No. of modular forms of weight $2k$.}
 	\label{tab:1}
 \end{table}

 \subsection{$\Gamma(3)$ modular group}
 It is a level three modular group which is isomorphic to discrete symmetry group $A_4$. The three modular forms of weight 2 present in this group can be expressed as:
 
 \begin{equation}
 \begin{aligned}
 & y_{1}(\tau)=\frac{i}{2\pi}[\frac{\eta'(\frac{\tau}{3})}{\eta(\frac{\tau}{3})}+\frac{\eta'(\frac{\tau+1}{3})}{\eta(\frac{\tau+1}{3})}+\frac{\eta'(\frac{\tau+2}{3})}{\eta(\frac{\tau+2}{3})}-27\frac{\eta'(3\tau)}{\eta(3\tau)}]\\
 &    y_{2}(\tau)=\frac{-i}{\pi}[\frac{\eta'(\frac{\tau}{3})}{\eta(\frac{\tau}{3})}+\omega^2\frac{\eta'(\frac{\tau+1}{3})}{\eta(\frac{\tau+1}{3})}+\omega\frac{\eta'(\frac{\tau+2}{3})}{\eta(\frac{\tau+2}{3})}]\\
 & y_{3}(\tau)=\frac{-i}{\pi}[\frac{\eta'(\frac{\tau}{3})}{\eta(\frac{\tau}{3})}+\omega\frac{\eta'(\frac{\tau+1}{3})}{\eta(\frac{\tau+1}{3})}+\omega^2\frac{\eta'(\frac{\tau+2}{3})}{\eta(\frac{\tau+2}{3})}]
 \end{aligned}
 \label{eqn:A6}
 \end{equation} 
 
  where $\eta(\tau)$ is the Dedekind eta-function and is defined in the following way:
  \begin{equation}
  \eta(\tau)=q^{\frac{1}{24}}\prod_{n=1}^{\infty}(1-q^{n}), \hspace{1cm} q=e^{2\pi i\tau}.
  \label{eqn:A7}
  \end{equation}
  
  The eta functions satisfy the equations \begin{equation}
  \eta(\tau +1)= \exp^{i\pi/12}\eta(\tau), \hspace{8mm} \eta(-1/\tau)=\sqrt{-i\tau} \eta(\tau)
  \label{eqn:A8}
  \end{equation}
   Another way of expanding these modular forms is the q-expansions, where q is expressed in terms of $\tau$ in the following way $q=\exp(2i\pi\tau)$. These expansions take the form:
   \begin{equation}
   \begin{aligned}
   & y_1(\tau)=1+12q+36q^2+12q^3+.........\\
   & y_2(\tau)=-6q^{1/3}(1+7q+8q^2+.......)\\
   & y_3(\tau)=-18q^{2/3}(1+2q+5q^2+......)\\
   \end{aligned}
   \label{eqn:A9}
   \end{equation}
   Moreover with the help of these lower weight modular forms we can construct the higher weight modular functions. For a detailed study one can refer to \cite{Zhang:2019ngf,Novichkov:2019sqv}.

\bibliography{file}	
\bibliographystyle{utphys}

	\end{document}